\documentclass[epj]{svjour}
\usepackage{mathrsfs,amssymb,amsmath,amsopn,bm,bbm,slashed,graphicx,color}

\def\eg{{\it e.g.}}

\def\ie{{\it i.e.}}

\newcommand{\beq}{\begin{equation}}
\newcommand{\eeq}{\end{equation}}
\newcommand{\bea}{\begin{eqnarray}}
\newcommand{\eea}{\end{eqnarray}}
\newcommand{\lsim}{\raisebox{-4pt}{$\,\stackrel{\textstyle <}{\sim}\,$}}

\newcommand{\fslash}{\slashed}
\newcommand{\ii}{\ensuremath{\mathrm{i}}}
\newcommand{\dd}{\ensuremath{\mathrm{d}}}
\newcommand{\e}{\ensuremath{\mathrm{e}}}
\newcommand{\D}{\ensuremath{\mathrm{D}}}

\newcommand{\GeV}{\ensuremath{\mathrm{GeV}}}

\newcommand{\fm}{\ensuremath{\mathrm{fm}}}
\newcommand{\diag}{\mathrm{diag}}
\newcommand{\erw}[1]{\ensuremath { %
    \left \langle {#1} \right \rangle}}

\DeclareMathOperator{\im}{Im}

\begin{document}
\title{Thermal Electromagnetic Radiation in Heavy-Ion Collisions}
\author{R. Rapp\inst{1}, H. van Hees\inst{2,3}
}
\institute{Cyclotron Institute and Department of Physics \& Astronomy,
Texas A{\&}M University, College Station, Texas 77843-3366, USA \and
Institut f{\"u}r Theoretische Physik, Goethe-Universit{\"a}t
  Frankfurt, Max-von-Laue-Str. 1, D-60438 Frankfurt, Germany \and 
  Frankfurt Institute of Advanced Studies (FIAS), Ruth-Moufang-Str. 1,
  D-60438 Frankfurt, Germany
}
\date{\today} 

\abstract{We review the potential of precise measurements of electromagnetic 
  probes in relativistic heavy-ion collisions for the theoretical understanding 
  of strongly interacting matter. The penetrating nature of photons and dileptons
  implies that they can carry undistorted information about the hot and dense
  regions of the fireballs formed in these reactions and thus provide a 
  unique opportunity to measure the electromagnetic spectral function of QCD
  matter as a function of both invariant mass and momentum. In particular we
  report on recent progress on how the medium modifications of the (dominant) 
  isovector part of the vector current correlator ($\rho$ channel) can shed 
  light on the mechanism of chiral symmetry restoration in the hot and/or dense
  environment. In addition, thermal dilepton radiation enables novel access to
  (a) the fireball lifetime through the dilepton yield in the low invariant-mass
  window $0.3 \; \GeV \leq M \leq 0.7 \; \GeV$,  and (b) the early temperatures 
  of the fireball through the slope of the invariant-mass spectrum in the 
  intermediate-mass region ($1.5 \; \GeV <M< 2.5 \; \GeV$). The investigation of
  the pertinent excitation function suggests that the beam energies provided by
  the NICA and FAIR projects are in a promising range for a potential discovery of 
  the onset of a first order phase transition, as signaled by a 
  non-monotonous behavior of both low-mass yields and temperature slopes.}

\maketitle
%
\section{Introduction}
Photons and dileptons have long been recognized as valuable probes of
the hot and dense medium created in heavy-ion collisions (HICs). Since
they can leave the reaction zone essentially unaffected by final-state
interactions, they mediate direct information on the properties of the
partonic and hadronic electromagnetic (EM) current-current correlation
function in the system in both time-like (dileptons) and light-like
(photons) domains. The thermal emission rates can be concisely written
in terms of the (retarded) spectral function of the EM
current-correlation function in the medium,
$\rho_{\rm em}=-{\rm Im}\Pi_{\rm em}/\pi$, as~\cite{MT84,gale-kap90}
\begin{alignat}{2}
\begin{split}
\label{Rll}
\frac{\dd N_{\ell \ell}}{\dd^4 x \dd^4 q} =&
-\frac{\alpha_{\text{em}}^2}{\pi^3 M^2} f_{\text{B}}(q_0;T) \\
& \times 
  \frac{1}{3} g_{\mu \nu} \im \Pi_{\text{em},\text{ret}}^{\mu
    \nu}(M,q;T,\mu_B)
\end{split}\\
\begin{split}
\label{Rgam}
q_0 \frac{\dd N_{\gamma}}{\dd^4 x \dd^3 \vec{q}} =&-\frac{\alpha_{\rm em}}{\pi^2} 
f_{\text{B}}(q_0;T) \\
&\times  \frac{1}{2} g_{\mu \nu} \im \Pi_{\text{em},\text{ret}}^{\mu
    \nu}(M=0,q;T,\mu_B) \ , 
\end{split}
\end{alignat} 
where $f_{\text{B}}$ denotes the thermal Bose distribution function and
$\alpha_{\text{em}} \simeq 1/137$ the EM coupling constant ($T$:
temperature, $\mu_B$: baryon chemical potential; $q^0$ and $q$ are the
energy and momentum of the dilepton or photon in the (local) rest frame
of the medium). The key difference between the two expressions is that
the dilepton rate, Eq.~(\ref{Rll}), depends on the invariant mass,
$M^2=q_0^2-q^2$, of the virtual photon, while the photon rate,
Eq.~(\ref{Rgam}), constitutes the $M\to0$ limit and thus only depends on
the photon's three-momentum (or energy), $q_0=q$. The rate expressions
are remarkably simple, as a product of a thermal distribution function
and the EM spectral function of the medium. Dilepton spectra are the
only known observables in HICs which provide direct access to a spectral
function of the QCD medium. This encodes a rich physics potential that
will be further elaborated below. In practice, the measured spectra
contain contributions from all stages of the evolution of a HIC,
requiring realistic bulk evolution models over which the rates need to
be integrated over.

\begin{figure}[!t]
\begin{center}
\includegraphics[width=0.95 \columnwidth]{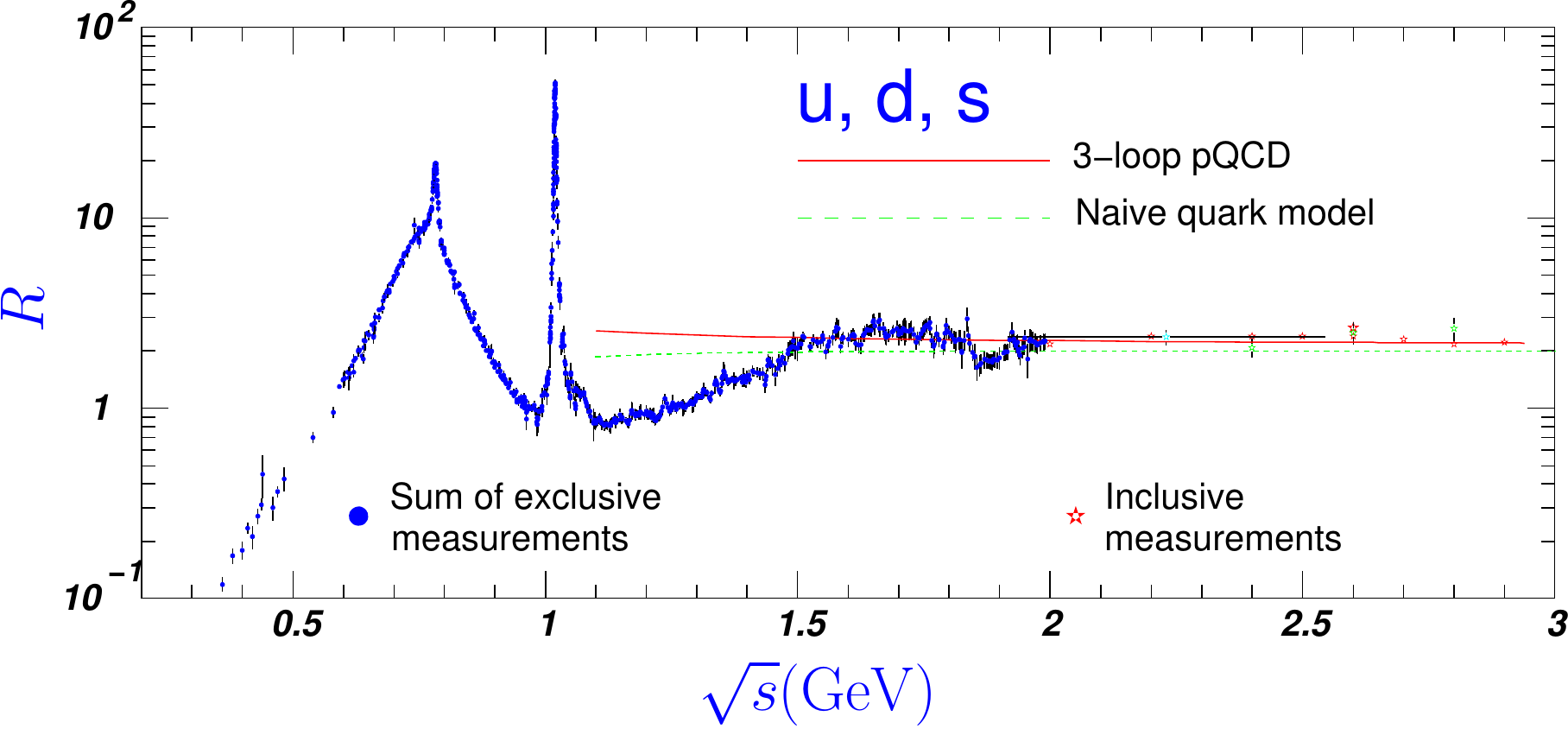}
\end{center}
\caption{Cross section ratio for $e^+e^-\to \text{hadrons}$ to
  $\e^+ \e^-\to \mu^+\mu^-$, which is directly proportional to the
  vacuum EM spectral function, $R=12\pi^2\rho_{\rm em}/M^2$.  Figure
  redrawn from Ref.~\cite{Nakamura:2010zzi}.}
\label{Rhad}
\end{figure}

The EM spectral function is well known in vacuum from the inverse
process of $\e^+\e^-$ annihilating into hadrons, cf.~Fig.~\ref{Rhad}. It
simply represents the excitation spectrum of the QCD vacuum in the
vector channel, $J^P=1^-$, \ie, for the quantum numbers of the
photon. For masses relevant for thermal radiation in HICs,
$M\lsim 3$\,GeV, it exhibits two basic regimes. In the low-mass region,
$M\lsim 1$\,GeV, the strength is concentrated in the light vector meson
peaks corresponding to $\rho$, $\omega$, and $\phi$. In the
intermediate-mass region, $1.5~{\rm GeV} \lsim M\lsim 3~{\rm GeV}$, the
strength of the hadronic spectrum is essentially given by the
perturbative $\text{q}\bar{\text{q}}$ continuum (subsequent
hadronization does not significantly affect the short distance processes
determining the cross section). The vacuum EM spectral function
therefore exhibits a clean transition from confined hadronic degrees of
freedom to weakly interacting quarks and anti-quarks as the resolution
($q^2=M^2$) of the probe is increased. In particular, the low-mass part
clearly signals the non-perturbative physics of the QCD vacuum related
to its quark and gluon condensate structures. By measuring dilepton
spectra in HICs one essentially puts this probe into the QCD medium.
This suggests that low-mass spectra can monitor the fate of hadrons as
temperature and density are increased and thus reveal underlying changes
in the condensates, often referred to as chiral symmetry restoration
associated with the melting of the quark-anti-quark condensate,
$\langle \bar qq\rangle$. As will be discussed below, most of the
low-mass radiation in high-energy HICs indeed emanates from temperatures
around the pseudo-critical chiral transition temperature,
$T_{\rm pc}^\chi\simeq 155$\,MeV \cite{Aoki:2006br,Bhattacharya:2014ara}, rendering
low-mass dileptons an excellent observable to investigate this
transition. On the other hand, for masses above $M\simeq 1.5 \; \GeV$,
temperature corrections to the EM spectral function are small (of order
${\cal O}(T^2/M^2$)), and dilepton spectra become an excellent
thermometer of the produced medium. As is well
known~\cite{Shuryak:1980tp,Rapp:2011is}, large invariant masses
exponentially favor the emission of early (hot) radiation, dominating
over the volume increase which grows with a power (typically 5-6) in
$T$.  While the temperature slope of the invariant-mass spectra involves
a certain average over a range of fireball temperatures, it is not
distorted by any (``Doppler'') blue shift due to the radial flow of the
expanding medium (invariant-mass distributions do not change under
Lorentz boosts). This is different for momentum spectra of thermal
photons (or dileptons), where an expanding medium imparts a significant
blue shift on the spectra which has to be deconvoluted before a ``true''
temperature can be extracted \cite{vanHees:2011vb}.

In the remainder of this contribution, we first briefly review the
relevance of chiral symmetry, its breaking and restoration in the
context of dileptons and then discuss recent work pertinent to it
(Sec.~\ref{sec:chiral}). We present the current state of phenomenology
of EM radiation in HICs and offer perspectives for the energy regime
covered by NICA (Sec.~\ref{sec:pheno}), followed by concluding remarks
(Sec.~\ref{sec:concl}).

\section{Chiral Symmetry and Dileptons}
\label{sec:chiral}
The QCD Lagrangian 
\begin{equation}
\begin{split}
\label{qcd}
\mathscr{L}_{\rm QCD} = \overline{q} (\ii \fslash{\D}-\hat{m}_q) q - \frac{1}{4}
G_{\mu \nu}^a G_{a}^{\mu \nu} \\
\text{with} \quad \D_{\mu}
=\partial_{\mu} + \ii g \frac{\lambda_a}{2} A_{\mu}^a
\end{split}
\end{equation}
is formulated in terms of quark fields $q$ carrying both color and
flavor indices with a diagonal current-quark mass matrix
$\hat{m}_q=\diag(m_u,m_d,m_s,\ldots)$. The gluon fields $A_{\mu}^a$ are
contracted with the 8 Gell-Mann matrices $\lambda_a$ acting in SU(3)
color space, with a non-Abelian field-strength tensor
$G_{\mu \nu}^{a} = \partial_{\mu} A_{\nu}^a - \partial_{\nu} A_{\mu}^a -
g f^{abc} A_{\mu}^b A_{\nu}^c$,
where the $f^{abc}$ are the SU(3) structure constants.  The chiral
symmetry of $\mathscr{L}_{\rm QCD}$ refers to its approximate invariance
under the spin-isospin rotations in the $u$ and $d$ sector,
\begin{equation}
\begin{split}
  q \rightarrow \exp(-\ii \vec{\alpha}_{\text{V}} \cdot \vec{\tau}/2) q \\
  \text{and} \quad q \rightarrow \exp(-\ii \gamma_5 \vec{\alpha}_{\text{A}} \cdot
  \vec{\tau}/2) q,
\end{split}
\end{equation}
where $\vec\tau$ are the SU(2) isospin matrices. This symmetry is
explicitly broken by small light quark masses,
$m_{u,d}\ll \Lambda_{\rm QCD}$ (which can be treated as a perturbation
giving rise to ``chiral perturbation theory'' as a low-energy effective
theory of QCD).  Chiral symmetry leads to (partially) conserved
isovector-vector and -axialvector currents,
\begin{equation}
\label{3a}
\vec{j}_V^{\mu} = \overline{q} \gamma^{\mu} \frac{\vec{\tau}}{2} q \ , \quad
\quad \vec{j}_A = \overline{q} \gamma^{\mu} \gamma_5 \frac{\vec{\tau}}{2} q \ ,
\end{equation}
also referred to as a chiral multiplet, usually associated with the
lowest-lying resonances $\rho(770)$ and $a_1(1260)$ in each
channel. Other chiral multiplets are identified as, \eg,
$\sigma(500)$-$\pi(140)$, $N(940)$-$N^*(1535)$.  The large splitting of
the chiral multiplets is commonly attributed to the spontaneous breaking
of chiral symmetry in the vacuum,
$\mathrm{SU}(2)_{\text{L}} \times \mathrm{SU}(2)_{\text{R}} \to
\mathrm{SU}(2)_{\text{V}}$,
induced by the formation of the quark condensate,
$\erw{\overline{q} q} \simeq 2 \fm^{-3}$ per quark flavor, where pions
arise as the pseudo-Goldstone modes of the unbroken directions of the
global symmetry.


\begin{figure*}[!t]
\begin{center}
\includegraphics[width=1.95 \columnwidth]{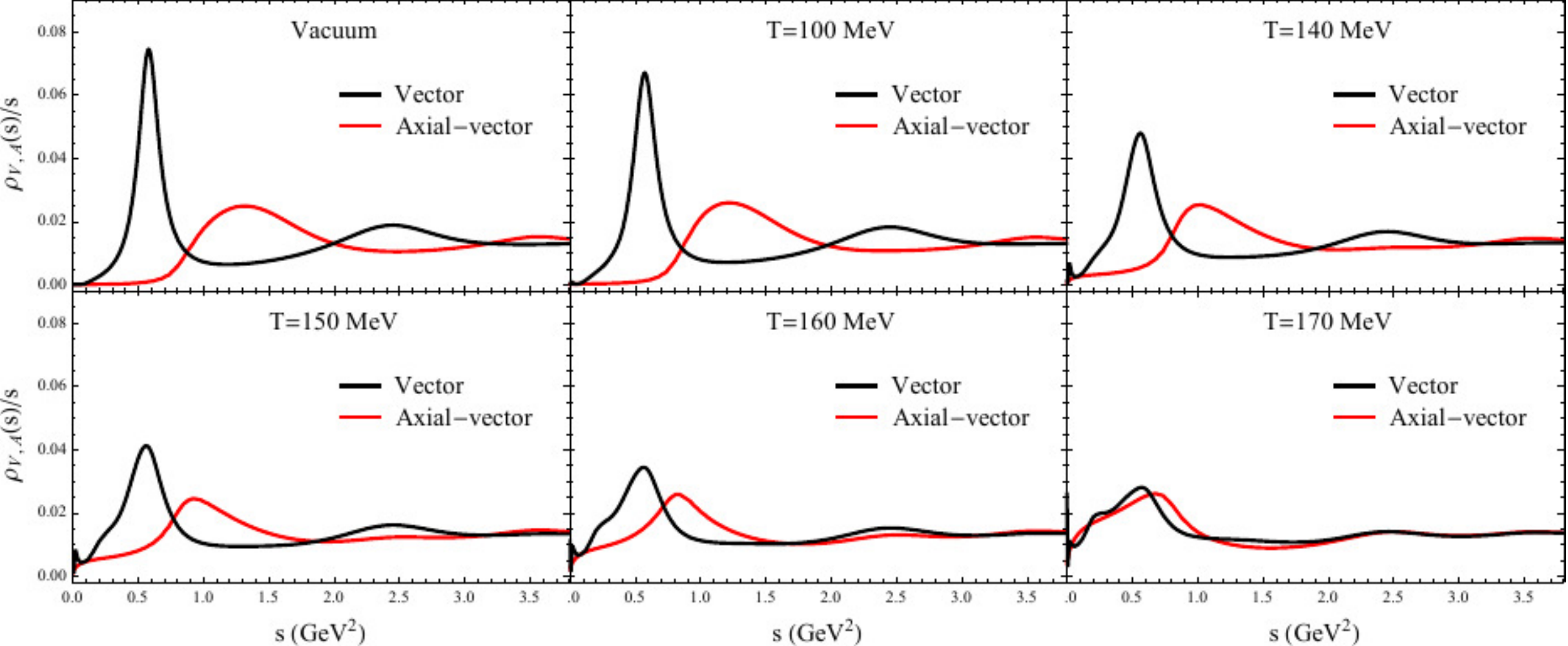}
\end{center}
\caption{Temperature progression (at $\mu_B=0$) of the axialvector
  spectral function (red lines) as inferred from a simultaneous solution
  of QCD and Weinberg sum rules~\cite{Hohler:2013eba} using as input an
  in-medium vector spectral function from hadronic many-body
  theory~\cite{Rapp:1999us} (black lines) and in-medium condensates as
  available from lattice QCD.}
\label{wsr}
\end{figure*}

As discussed above, the low-mass part of the EM spectral function is well 
described by the spectral functions of the light vector mesons, 
\begin{equation}
\label{rho-em}
\rho_{\text{em}} = -\frac{1}{\pi} \im \left[\frac{m_\rho^4}{g_\rho^2}D_{\rho} + 
\frac{m_\omega^4}{g_\omega^2}  D_{\omega} + \frac{m_\phi^4}{g_\phi^2} D_{\phi} \right ],
\end{equation}
known as the vector-meson dominance (VMD)~\cite{gounaris:1968,klz67}
hypothesis, which arises from the current-field identity
$j_V^\mu=(m_V^2/g_V) V^\mu$ ($V \in \{\rho,\omega,\phi \}$).
Since the contributions to $\rho_{\rm em}$ are dominated by the $\rho$
meson, calculations of medium effects have focused on the
latter. Hadronic many-body approaches generically find a strong
broadening of the $\rho$-meson spectral function in hot/dense matter,
leading to a melting of its resonance structure mostly driven by
interactions with baryons and anti-baryons~\cite{Rapp:1999ej}. These
approaches lead to a good description of all available dilepton data in
HICs (see Sec.~\ref{sec:pheno} below). A rigorous relation of the $\rho$
melting to chiral restoration has not been established yet, but progress
has been made recently.

\begin{figure}[!t]
\centerline{\includegraphics[width=0.79\columnwidth]{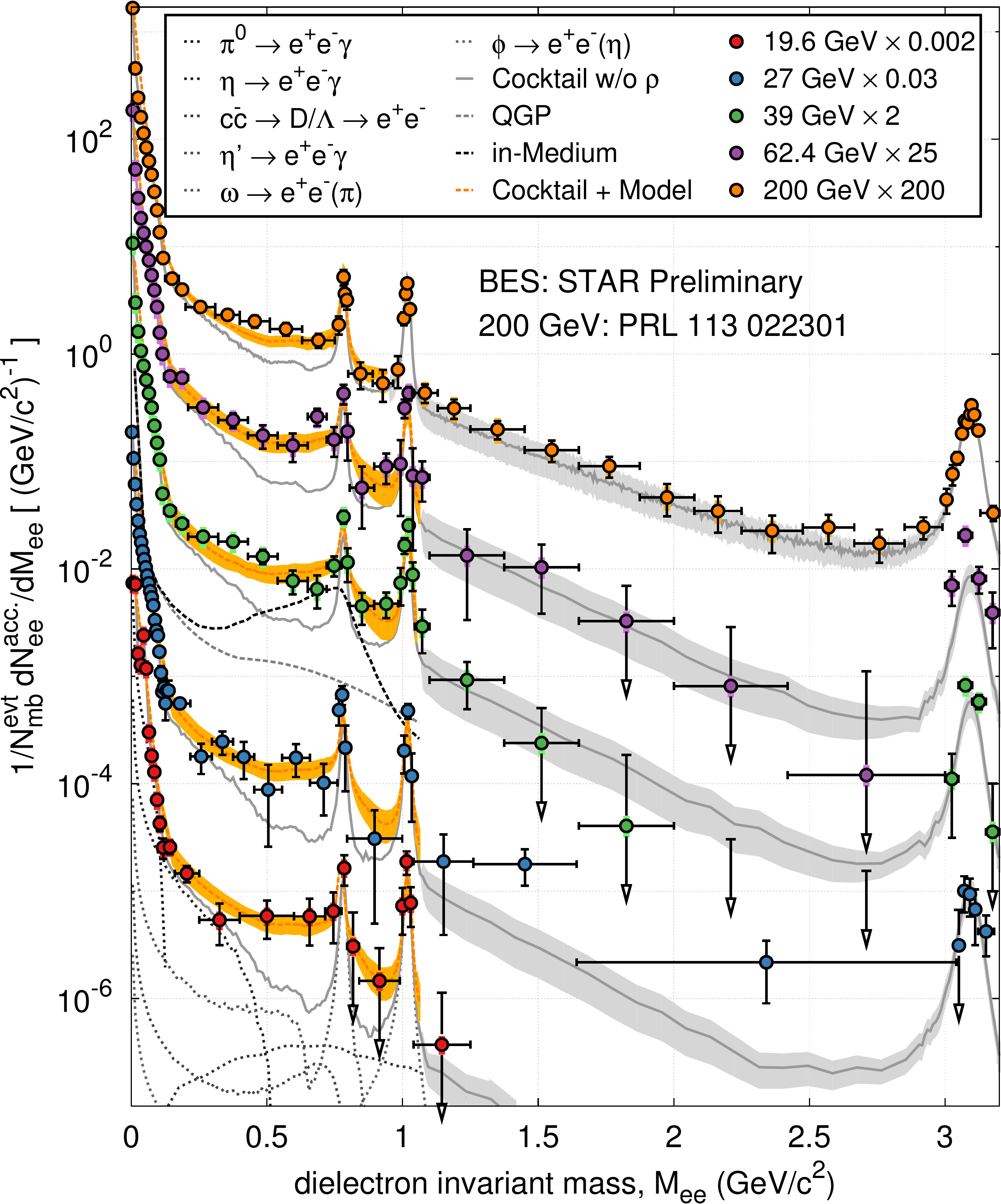}}

\vspace{0.5cm}

\centerline{\includegraphics[width=0.75 \columnwidth]{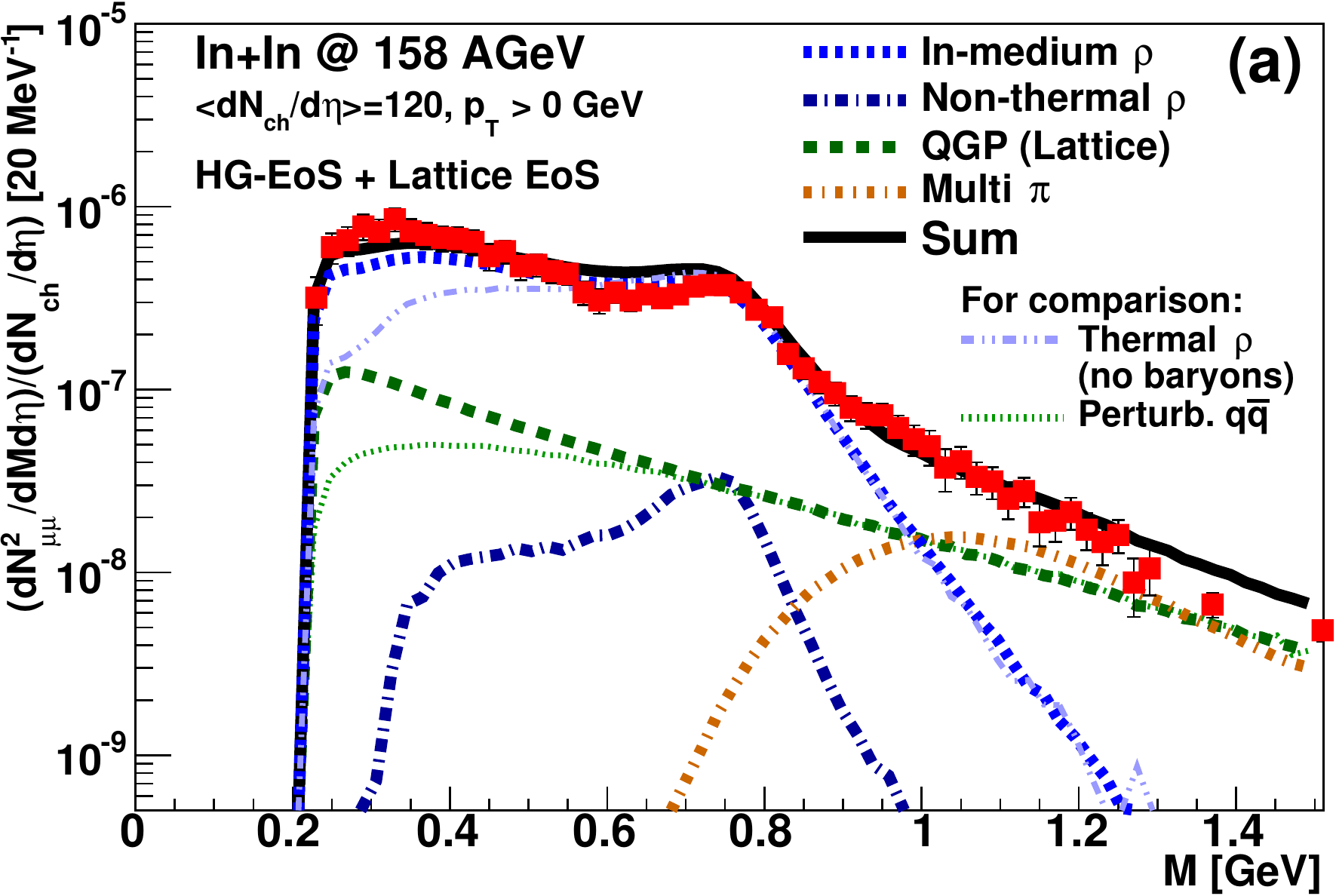}}

\vspace{0.5cm}

\centerline{\includegraphics[width=0.72 \columnwidth]{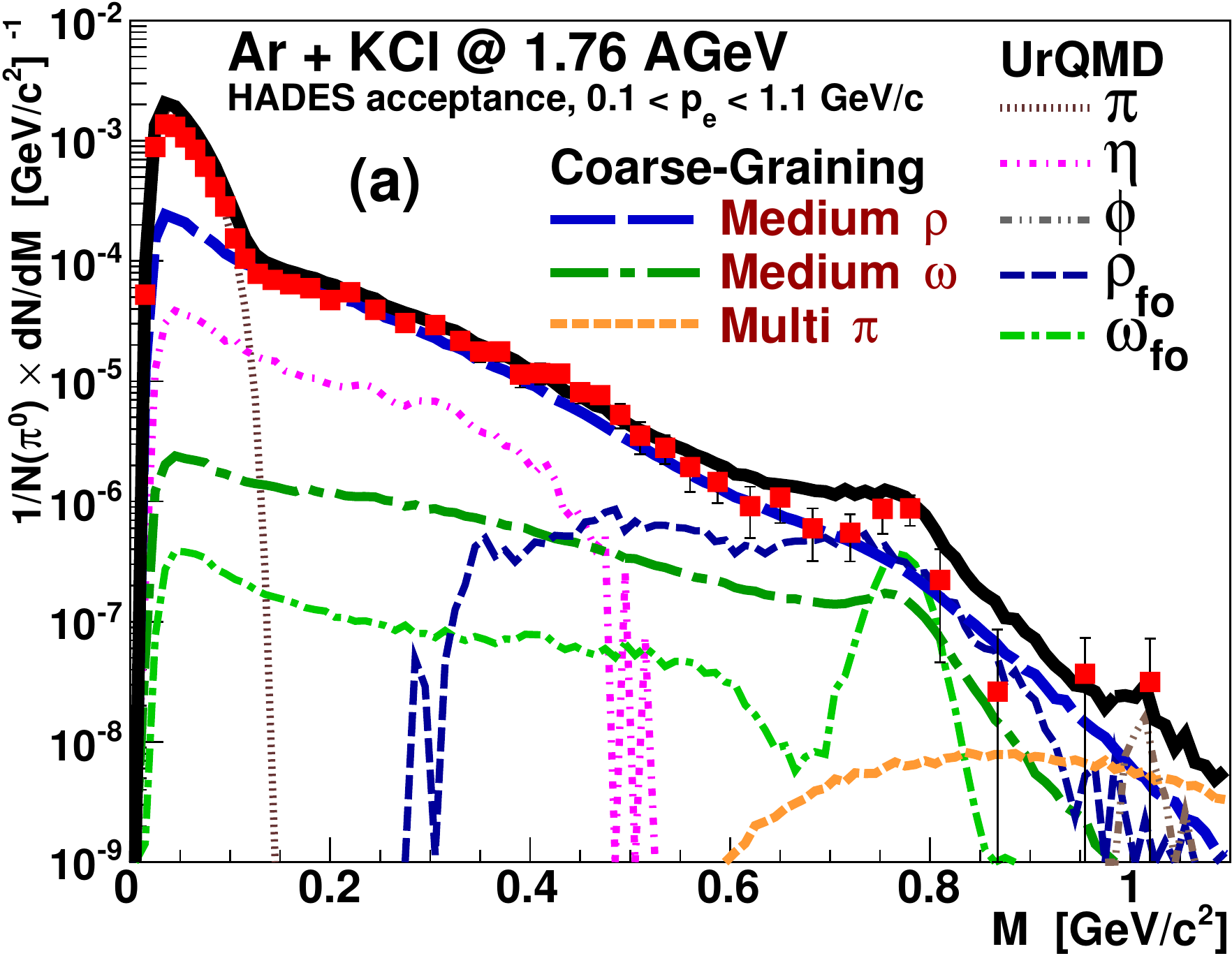}}

\vspace{0.2cm}

\caption{Dilepton invariant-mass spectra from SIS-18 to top RHIC
  energies using the in-medium $\rho$ spectral function of
  Ref.~\cite{Rapp:1999us} for hadronic emission and in-medium $q\bar q$
  annihilation from the QGP. Top panel: excitation function measured in
  Au-Au collisions by STAR~\cite{Huck:2014mfa,Adamczyk:2015lme},
  compared to predictions using a thermal fireball
  model~\cite{Rapp:2013nxa} (Figure reprinted with permission from
  \cite{Huck:2014mfa}).  Lower two panels: NA60 data in In-In at
  CERN-SPS (middle panel) and Ar-KCl at GSI-SIS-18 (lower panel)
  calculated with the same dilepton emission rates implemented into a
  coarse-grained UrQMD approach~\cite{Endres:2014zua,Endres:2015fna}.}
\label{dilep}
\end{figure}

In Ref.~\cite{Hohler:2013eba}, a combined analysis of finite-temperature
QCD~\cite{Hatsuda:1992bv} and Weinberg~\cite{Kapusta:1993hq} sum rules
has been carried out to test the in-medium $\rho$ spectral function that
describes dilepton spectra~\cite{Rapp:1999us} with respect to chiral
restoration. QCD sum rules relate power series with quark and gluon
condensates to spectral functions for a given quantum number, while
Weinberg sum rules specifically relate moments of the difference between
isovector-vector ($\rho$) and -axialvector ($a_1$) spectral functions to
chiral order parameters (pion decay constant, quark condensate,
etc.). Using the latter as input from lattice-QCD (augmented by hadron
resonance gas predictions where needed) together with calculated
in-medium $\rho$ spectral functions, viable in-medium $a_1$ spectral
functions were searched for that can satisfy both QCD and Weinberg sum
rules within their typical accuracy of $\sim 0.5\%$. A solution was
found that gradually degenerates with the vector-meson spectral
function, cf.~Fig.~\ref{wsr}. This implies that our current
understanding of dilepton data is compatible with (the approach toward)
chiral symmetry restoration. This analysis also suggests a mechanism of
chiral restoration by which the broadening of both $\rho$ and $a_1$ is
accompanied by a reduction of the $a_1$ mass moving toward the $\rho$
mass, while the latter approximately stays constant.  A microscopic
investigation of the in-medium $\pi$-$\rho$-$a_1$ system has recently
been conducted within the Massive-Yang Mills (MYM)
framework~\cite{Hohler:2015iba}. 
First, the notorious difficulties of
the MYM approach to describe the vacuum axialvector spectral function
could be overcome by introducing a broad $\rho$ propagator into the
$a_1$ self-energy, accompanied by judiciously chosen vertex corrections
to maintain PCAC~\cite{Hohler:2013ena}.  A one-loop finite-$T$
calculation results in a moderate broadening of both peaks, as well as a
downward mass shift of the $a_1$ by up to about 200\,MeV along with a
15-20\% reduction of the pion decay constant and scalar condensate. This
result is remarkably similar to the phenomenological sum rule analysis
and corroborates the ``burning'' of the chiral mass splitting as a
mechanism of chiral degeneracy. The extension of this analysis to finite
baryon densities remains a formidable task. Finally, a recent
lattice-QCD calculation has investigated the behavior of the nucleon
correlation function and its chiral partner, $N^*(1535)$, at finite
$T$~\cite{Aarts:2015mma}. Also here it was found that the mass of the
ground state (nucleon) remains essentially stable while the mass of the
excited state approaches the former and degenerates with it around
$T_{\rm pc}$.

\section{Dilepton Production in Heavy-Ion Collisions}
\label{sec:pheno}

Large efforts have been devoted over the past decades to calculate
photon and dilepton spectra in HICs in comparison to experimental
data. For low-mass dilepton spectra, a consistent picture has emerged in
terms of a broadening $\rho$ spectral function in the hadronic phase
which smoothly merges into a quark-antiquark continuum in the QGP around
temperatures of $T_{\rm pc}\simeq170$\,MeV.  When implemented into an
expanding thermal fireball which utilizes a modern lattice-QCD based
equation of state (EoS) and is tuned to fit hadronic yields and spectra,
the high-precision NA60 data~\cite{Specht:2010xu} in In-In collisions
(as well as NA45/CERES data~\cite{Adamova:2002kf,Agakichiev:2005ai} in
Pb-Au collisions) at CERN-SPS energies ($\sqrt{s}$=17.3\,GeV) are well
described; pertinent predictions for the excitation function in Au-Au
collisions for $\sqrt{s}$=19.6, 27, 39, 62.4 and 200\,GeV agree with the
STAR data~\cite{Huck:2014mfa,Adamczyk:2015lme}, cf.~upper panel in
Fig.~\ref{dilep}. Notably, the large enhancement reported earlier by
PHENIX~\cite{Adare:2009qk} has recently been revisited: their most
recent data~\cite{Adare:2015ila} are now consistent with the STAR data
and theory predictions.  At lower energies, larger initial overlap times
of the colliding nuclei as well as prolonged thermalization times may
render bulk evolution models based on rapid local equilibration
(fireball and/or hydrodynamics) problematic. In order to still utilize
the conceptually sound concept of thermal EM emission rates, a
coarse-graining approach of a realistic bulk medium evolution has
recently been implemented~\cite{Endres:2015fna,Endres:2014zua}:
well-tested UrQMD transport simulations are averaged over many events on
a space-time grid, where the local baryon densities and $p_t$ spectra
are mapped to a hadron resonance gas EoS for $T\le T_{\rm pc}$ and a
lattice-QCD based EoS above.  Using the same dilepton rates as in the
fireball calculations, the NA60 data are reproduced with similar quality
as in the fireball model~\cite{Endres:2014zua}, cf.~middle panel of
Fig.~\ref{dilep}.  Running the same approach at the much lower energies
of SIS-18 at GSI, the HADES data are remarkably well reproduced without
free parameters, cf.~lower panel of Fig.~\ref{dilep}.

\begin{figure}[!t]
\centerline{\includegraphics[width=0.85 \columnwidth]{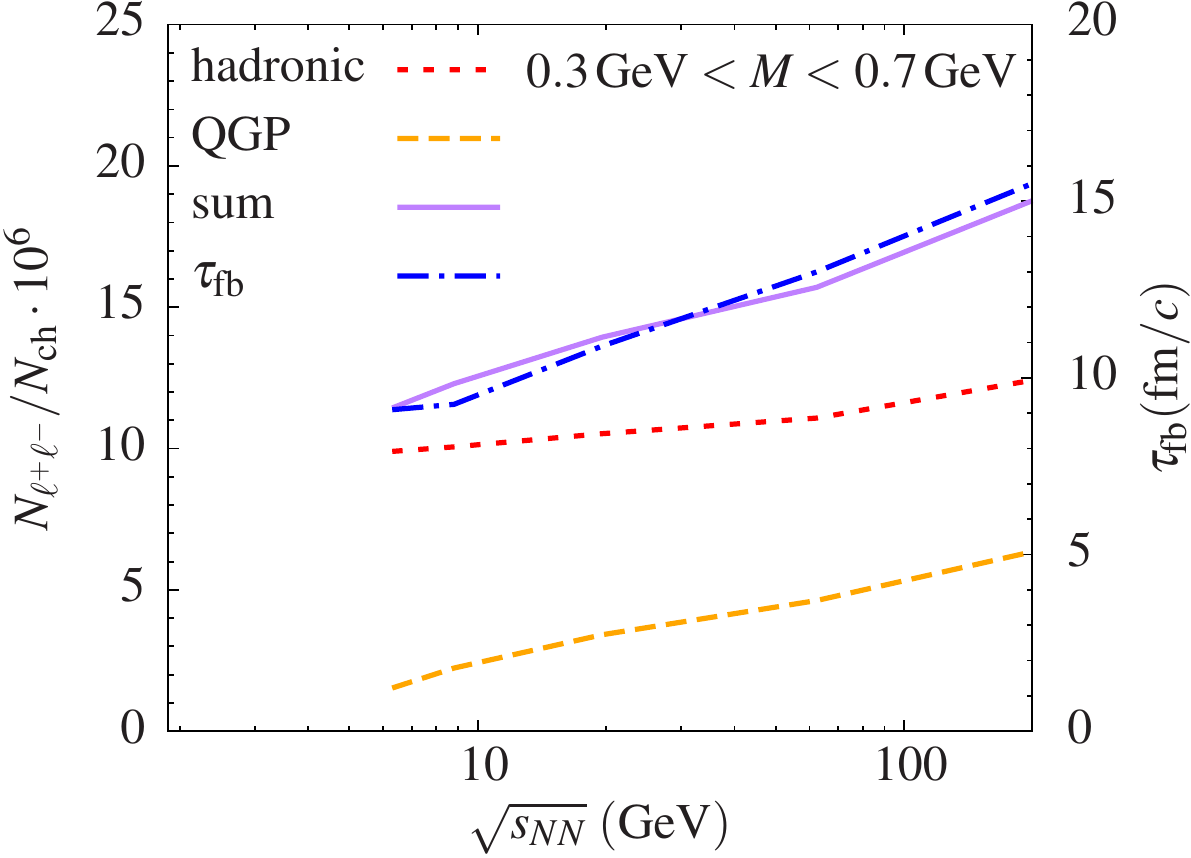}}

\vspace{0.3cm}

\centerline{\includegraphics[width=0.79 \columnwidth]{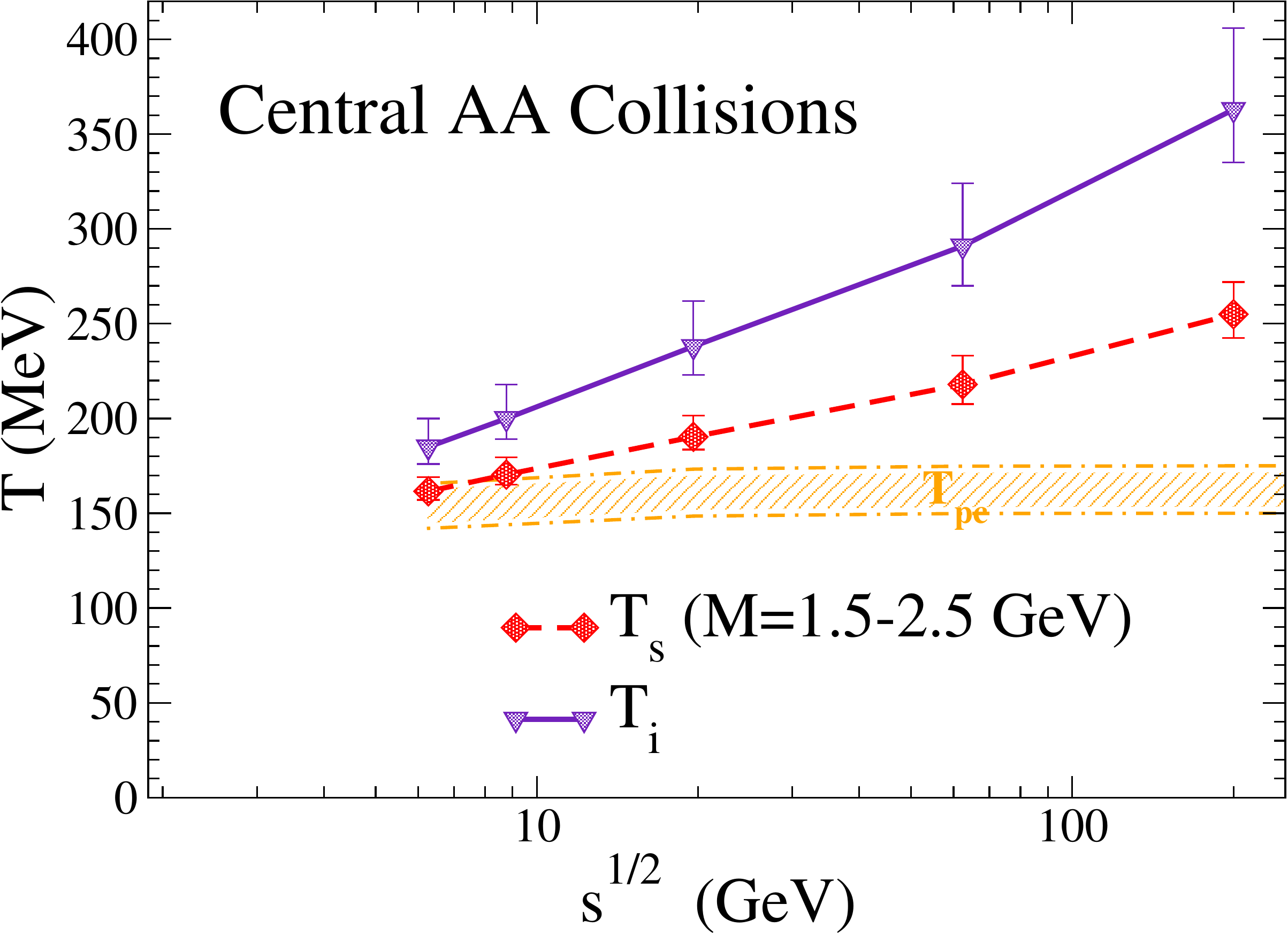}}

\vspace{0.2cm}

\caption{Excitation function of thermal dilepton radiation in central
  HICs. Upper panel: The integrated dilepton yield in the mass range
  $0.3 \;\GeV \leq M_{\text{ee}} \leq 0.7 \; \GeV$, serving as fireball
  chronometer; lower panel: The inverse slope parameter over the
  invariant-mass range
  $1.5 \; \GeV \leq M_{\text{ee}} \leq 2.5 \; \GeV$, serving as fireball
  thermometer. Figures taken from Ref.~\cite{Rapp:2014hha}.}
\label{ex-fct}
\end{figure}
The robust understanding of all existing low-mass dilepton spectra
allows to take this observable to the next level, by exploiting it as a
probe of fundamental fireball properties across the QCD phase diagram.
In Ref.~\cite{Rapp:2014hha} this has been put forward in two respects,
by extracting (a) the total fireball lifetime from the excess yields in
the low-mass region, and (b) the early fireball temperatures from the
invariant-mass slopes in the intermediate-mass region.  The results for
central AA collisions (with atomic number $A \simeq 200$) in the
ultrarelativistic energy regime, $\sqrt{s}>6 \; \GeV$, are summarized in
Fig.~\ref{ex-fct}. The upper panel shows that the thermal dilepton yield
in the mass range $0.3\; \GeV \leq M_{\text{ee}} \leq 0.7 \; \GeV$ is an
excellent measure of the total lifetime of the fireball (other mass
regions, \eg, around the free $\rho$ mass do not work as well); this
could become particularly useful if lifetime variations occur due to a
softening of the EoS, possibly resulting in a non-monotonous energy
dependence. The lower panel shows that the slope of the spectrum
extracted over the range
$1.5 \; \GeV \leq M_{\text{ee}} \leq 2.5 \;\GeV$ is a direct measure of
the temperature in the early phases of the fireball evolution, well
inside the QGP at top-SPS energy and beyond, but reflecting and tracing
the mixed phase for $\sqrt{s}\lsim10 \; \GeV$. A possible plateau toward
lower energies, before dropping to the lower temperatures of the SIS-18
regime, would be a rather strong evidence for the onset of a mixed phase
and thus the discovery of a first-order transition.

\section{Conclusions}
\label{sec:concl}
Dileptons remain a prime probe of the interior medium of the fireballs
produced in HICs.  A robust understanding of existing data has emerged
over the last decade, in terms of a melting of the $\rho$ meson as the
phase boundary is approached from below. This is indicative of a change
from hadronic to partonic degrees of freedom and compatible with the
restoration of the spontaneously broken chiral symmetry. While the
precise mechanism for the restoration remains to be determined, current
theoretical models suggest the degeneracy of chiral multiplets to be
realized by burning off the chiral mass splitting in the vacuum with the
ground state mass being stable. We have shown how thermal dilepton
radiation can be taken to the next level as a probe of fundamental
fireball properties such as its total lifetime and early temperatures. A
particularly promising regime for the pertinent excitation functions
turns out to be collision energies of $\sqrt{s}\simeq 2$-$8\; \GeV$.
Since this regime is uncharted territory in terms of low-mass dilepton
measurements with heavy ions to date, its exploration at the future NICA
facility is a compelling case. The situation might become a bit
more involved if novel spectral structures emerge in the vector
spectral function close to the putative critical point. However, this
rather provides an additional potential for new discoveries.
\\

\noindent
{\bf Acknowledgments} \\
This work has been supported by the US-NSF under grant No.~PHY-1306359.



\begin{flushleft}

\end{flushleft}

\end{document}